\documentclass[useAMS]{mn2e}

\usepackage{amsmath,amsfonts}
\usepackage{epsfig}

\def\reff@jnl#1{{\rm#1\/}}

\def\aj{\reff@jnl{AJ}}                  
\def\araa{\reff@jnl{ARA\&A}}            
\def\apj{\reff@jnl{ApJ}}                        
\def\apjl{\reff@jnl{ApJ}}               
\def\apjs{\reff@jnl{ApJS}}              
\def\ao{\reff@jnl{Appl.Optics}}         
\def\apss{\reff@jnl{Ap\&SS}}            
\def\aap{\reff@jnl{A\&A}}               
\def\aapr{\reff@jnl{A\&A~Rev.}}         
\def\aaps{\reff@jnl{A\&AS}}             
\def\azh{\reff@jnl{AZh}}                        
\def\baas{\reff@jnl{BAAS}}              
\def\jrasc{\reff@jnl{JRASC}}            
\def\memras{\reff@jnl{MmRAS}}           
\def\mnras{\reff@jnl{MNRAS}}            
\def\pra{\reff@jnl{Phys.Rev.A}}         
\def\prb{\reff@jnl{Phys.Rev.B}}         
\def\prc{\reff@jnl{Phys.Rev.C}}         
\def\prd{\reff@jnl{Phys.Rev.D}}         
\def\prl{\reff@jnl{Phys.Rev.Lett}}      
\def\pasp{\reff@jnl{PASP}}              
\def\qjras{\reff@jnl{QJRAS}}            
\def\skytel{\reff@jnl{S\&T}}            
\def\solphys{\reff@jnl{Solar~Phys.}}    
\def\sovast{\reff@jnl{Soviet~Ast.}}     
\def\ssr{\reff@jnl{Space~Sci.Rev.}}     
\def\zap{\reff@jnl{ZAp}}                        
\def\nat{\reff@jnl{Nature}}             

\title[Interstellar C$_2$  absorption lines toward  Cernis 52 ]{Interstellar C$_2$  in the Perseus molecular complex: excitation temperature and density of a molecular cloud with anomalous microwave emission.}

\author[S. Iglesias-Groth] {Susana Iglesias-Groth$^{1,2}\thanks{E-mail: sigroth@iac.es}$ \\ 
$^1$ Instituto de Astrofis\'{\i}ca de Canarias, 38200 La Laguna, Tenerife, Canary Islands, Spain \\
$^2$ Universidad de La Laguna, E-38205 La Laguna, Tenerife, Spain \\
}

\date{Accepted Received In original form}
\pagerange{\pageref{firstpage}--\pageref{lastpage}}
\pubyear{2010}

\begin{document}

\label{firstpage}
\maketitle

\begin{abstract}
Interstellar  absorption lines up to J"=10 in the (2,0) band  and up to J"=6 in the (3,0) band of the C$_2$ $A^1\Pi_u$ - $X^1\Sigma^+_g$ system are detected toward star Cernis 52 (BD+31$^o$ 640) in the Perseus
molecular complex. The star lies   in a redenned  line of sight    where various experiments have  detected anomalous microwave emission spatially correlated with dust thermal emission.  The inferred total C$_2$ column density of N(C$_{2}$) = (10.5$\pm$ 0.2) x  10$^{13}$ cm$^{-2}$ is  well correlated with that of CH as expected from
 theoretical models and is among the highest reported on translucent clouds with similar extinction.  The observed rotational C$_2$ lines  constrain the gas-kinetic temperature T and the density n=n(H)+n(H$_2$) of
 the intervening cloud to T = 40$\pm$10 K and n = 250 $\pm$ 50 cm$^{-3}$, respectively.  This is the first determination of gas-kinetic temperature and particle density of a  cloud with known  anomalous microwave emission.

\end{abstract}

\begin{keywords}
ISM:molecules---ISM:lines and bands---ISM:abundances
\end{keywords}

\section{Introduction}
 
The  moderately reddened [E(B-V)=0.9] early A-type star Cernis 52 (R.A 03 43 01; Dec +31 58 10; Cernis 1993 ) lies in the line of sight of the Perseus molecular complex which is located at a distance of about 240 pc.  The star is  a likely member of the very young IC 348 cluster, about 1 degree in projection away from its core (Gonz\'anlez-Hern\'andez et al. 2009). This line of sight towards Perseus is remarkable because of  the anomalous microwave emission detected by Watson et al (2005) in the frequency  range 10-60 GHz. According to recent data,  this  anomalous emission  reaches a maximum in the line of sight of Cernis 52 (Tibbs et al.~2010).  This new microwave emission process  cannot be explained by classical  mechanisms as synchrotron, free-free or thermal emission from dust particles, however it is  spatially  correlated with the distribution of dust  traced by IRAS images. 

Draine and Lazarian (1998)  postulated that the anomalous microwave emission could be due to  electric dipole radiation of  rapidly spinning small interstellar carbon based molecules, as for example   polycyclic aromatic hydrocarbons (PAHs) (see also Iglesias-Groth 2005 for the potential contribution of hydrogenated forms of fullerenes).  The recent detection of optical bands in the spectrum of Cernis 52 which are consistent with  transitions of the  naphthalene C$_{10}$H$_8$$^+$ and anthracene C$_{14}$H$_{10}$$^+$  cations (Iglesias-Groth et al 2008, 2010) add support to this PAH hypothesis and call for a more extensive study of the physical and chemical  conditions of the intervening material. Interstellar CH, CH$^+$ have also  been detected  toward Cernis 52 with high column densities  suggesting that  the H$_2$ column density  in the intervening cloud is also high (Iglesias-Groth et al. 2010). CH  is empirically correlated with C$_2$  in diffuse and translucent molecular clouds (see e.g. van Dishoeck \& Black 1989, Federman et al. 1994, Gredel 1999) thus a high column density of C$_2$ in the intervening cloud  may be expected. The  basic chemical processes which lead to the formation of diatomic carbon bearing species were reviewed by Federman \& Huntress (1989) and the dominant C$_2$ formation path is C$^+$ + CH $\rightarrow$  C$_2$$^+$ + H.

We report here the detection of  several  interstellar absorption lines of the simplest multicarbon molecule  C$_2$ toward Cernis 52 and  the study of physical parameters such as gas-kinetic temperatures and particle densities in the intervening cloud. We identify many weak absorption lines  of C$_2$   leading to the estimation of column densities for levels $J^{"}$ $<$ 12. The theory of C$_2$ excitation developed  by Chaffee et al. (1980) and  van Dishoeck \& Black (1982) has been  considered by a variety of authors in their analysis of interstellar C$_2$ lines  (see e.g. Hobbs 1981, Gredel 1999, Gredel et al 2001, Sonnentrucker et al. 2007, Ka\'zmierczak et al. 2010)  to derive densities and temperatures in diffuse and translucent molecular clouds.  We  use the van Dishoeck \& Black  formalism to infer the C$_2$ total abundance,  the kinetic temperature and  the density of the gas in the intervening cloud. We also report the detection of an  absorption feature in the spectrum of Cernis 52 which can be abscribed to the   4051.6 \AA~ band of C$_3$.

\begin{figure*}
\includegraphics[angle=0,width=11cm,height=11cm]{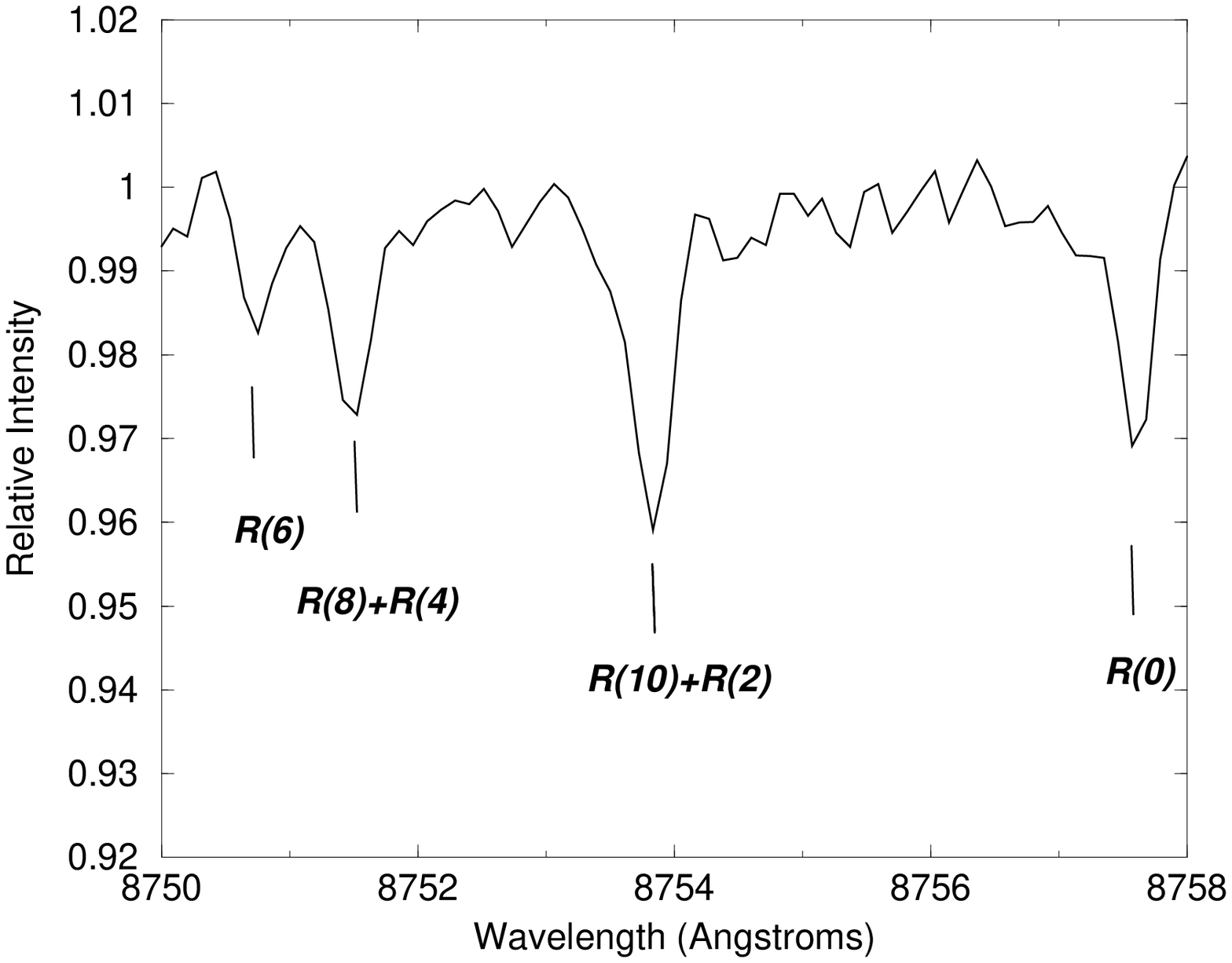}
\caption{Fig. 1. Spectrum covering the (2,0) band of the C2 Phillips System towards Cenis 52. Detected rotational lines are 
identified. }
\label{fig:f1}
\end{figure*}

\section{Observations and data reduction}

Interstellar absorption lines which arise from the (2,0) and (3,0) bands of the C$_2$ Phillips system around 8765 abd 7720 \AA, respectively, were searched toward star Cernis 52. The observations were  carried out   in  November 2008  using  the High Resolution Spectrograph (HRS) of the 9.2m Hobby-Eberly Telescope (HET) at McDonald Observatory (Texas, USA). Exposures of the star's spectrum were obtained on each of four nights (November 4, 7, 9  and 29) at a resolving power  R=40000. In all, 15 exposures lasting 20 minutes each were obtained and were subsequently combined during data reduction.  Fast  rotating  stars were also observed with the same instrument on the same nights to allow  removal  of telluric lines and correction of the instrumental response.  
The data were reduced using IRAF and  wavelength calibrated using ThAr lamps. The dispersion of the re-binned linear data was 96 m\AA/pixel. The fiber used to feed  HRS  led to a spectral resolution of 0.25 \AA~ in the spectral regions of interest.  The accuracy of the wavelength calibration of each order was better than 10 m\AA~ as shown by measurements  of the wavelengths of telluric lines recorded in various echelle orders. After wavelength calibration, the   Cernis 52 spectra  were  corrected for telluric line contamination and possible instrumental effects dividing each individual spectrum  by the featureless spectrum of  a much  brighter, hot and fast rotating star in a nearby line of sight observed with the same instrument configuration. Individual spectra  where then combined to improve S/N. The final spectrum  achieved S/N $\geq$ 300 per pixel.   

 Cernis 52 is embedded in a cloud responsible for significant visible extinction  and the presence in the spectrum of  molecular absorption  bands  caused by the intervening interstellar material is therefore  expected.  Figures 1 and 2 show interstellar absorption lines which arise from the (2,0) and (3,0) bands of the C$_2$ $A^1$$\Pi_u$ - $X^1$$\Sigma^+$$_g$ Phillips System, around 8765 \AA~ and 7720 \AA,  respectively. We identified and measured absorption lines (P,Q and R branches) in  bands (2,0) 8750-8849 \AA~ and some of (3,0) 7714-7793 \AA. The equivalent widths in the final normalized  spectra were measured by fitting a Gaussian profile to each absorption line using the IRAF task SPLOT.  In the case of unresolved line blends, such as the (2,0) R(10)+R(2) blend,  we used the deblending option offered by the same routine to  determine the equivalent width of each component. The equivalent widths with errors for all the measured interstellar lines of C$_2$ are listed in Table 1. The observational  uncertainties in the equivalent widths  are typically $\pm$1 m\AA, except for those lines marked as blends which have errors of $\sim$ 2 m\AA.  The table lists rest wavelengths in air  obtained from the wavenumbers in vacuum of Chauville, Maillard and Mantz (1977) and converted to air wavelengths using Birch \& Downs (1993).

\label{fig:f2}
\begin{figure*}
\includegraphics[angle=0,width=10cm,height=10cm]{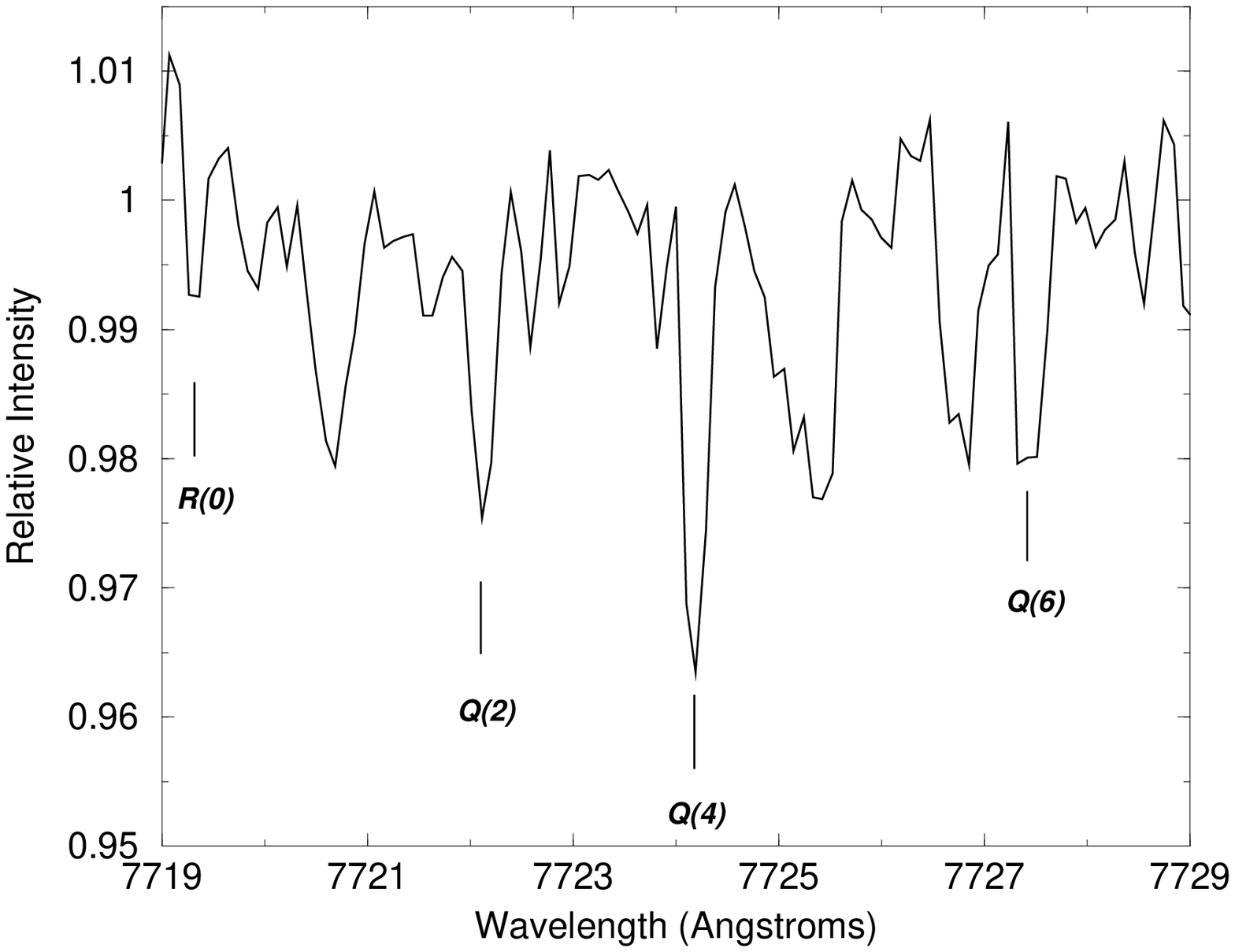}
\caption{ Spectrum covering the (3,0) band of the C2 Phillips System towards Cernis 52. Detected rotational lines are
identified}
\label{fig:f2}
\end{figure*} 

\section{Results and Analysis}

\subsection{Molecular parameters}

The measured equivalent widths W$_{\lambda}$ of the C$_2$ lines were converted into column densities for  individual  rotational levels N(J")  assuming that the absorption lines are optically thin  (i.e. that the absorption lines are on the linear part of the curve of growth). This approximation is accurate to better than 10 per cent for W$_{\lambda} <$ 10 m\AA~ and Doppler widths b $>$ 1 km s$^{-1}$ (Stromgren 1948).  We used the following  equation with W$_{\lambda}$   in units of \AA~ and N in units of cm$^{-2}$ 
\begin{equation}
N(J^{"}) = 1.13x10^{20} W_{\lambda}/(f_{J^{'}J^{"}} \lambda^2) 
\end{equation}
where  $\lambda$ is wavelength and f$_{J^{'}J^{"}}$ is the absorption oscillator strength.   

Line oscillator strengths f$_{J^{'}J^{"}}$ were calculated from the (2,0) band oscillator strength f$_{20}$ from the relation f$_{J^{'}J^{"}}$ =f$_{20}$ $(\nu_{J^{'}J^{"}}$/$\nu_{band}$)    S$_{J^{'}J^{"}}$/(2(2$J^{"}$ + 1)),   and
H\"onl-London rotational line intensity factors S$_{J^{'}J^{"}}$ of ($J^{"}$ +2), (2$J^{"}$ +1), and ($J^{"}$ - 1) for the R, Q, and P lines, respectively. The general formulae for the H\"onl-London factors (Herzberg 1950) are
simplified to the former values and normalized such that for each J" the sum $\Sigma_{J^{'}}$ S$_{J^{'}J^{"}}$/ g (2 J$^{"}$+1) = 1,  where g=2  is the value of the  ratio of electronic degeneracy factor  for the C$_2$ A-X system.   
Here $\nu_{band}$  is the wavenumber of the band, equal to 11413.91 cm$^{-1}$ for the Phillips (2-0) band. For the absorption oscillator strength of the (2-0) Phillips band we used the value  f$_{20}$ = 1.36 ($\pm$0.15) x 10$^{-3}$
measured by Erman and Iwamae (1995) which compares well with 1.44 x 10$^{-3}$ as  obtained from an ab initio calculation by van Dishoeck (1983) and with  the value  f$_{20}$ = 1.2 x 10$^{-3}$  suggested by Lambert et al. (1995). We infer 
the band oscillator strength for the (3,0) band from f$_{20}$ using the theoretical ratio of f$_{20}$/f$_{30}$ = 2.2 (van Dishoeck \&  Black 1982).


 \subsubsection{Column densities and Rotational excitation of C$_2$}

As discussed  by Gredel et al. (2001) a curve of growth analysis shows that for a typical value of b= 1 km s$^{-1}$  the C$_2$  lines suffer from saturation for W$_{\lambda}$ $>$ 15 m\AA. All our C$_2$ lines are weaker than this value and therefore we have ignored saturation corrections.  The derived column densities for each rotational level  N (J") are listed in Table 1 with their  uncertainties. An independent check on the reliability of the results can be made by a comparison of the column densities derived from the P, Q and R lines which arise from the same lower level $J^{"}$. Because of the smaller oscillator strength of P lines they proved very difficult to measure. Thus, the  empirical uncertainties in N(J")  of order 20 \%  are obtained by comparing for each J" the values obtained from R(J) and Q(J) lines.

\begin{table*}
\begin{minipage}{170mm}
\begin{center}
\caption{Summary of C$_2$ lines toward Cenis 52. (a-blend line).
\label{tab:table_1}}
\scriptsize{
\begin{tabular}{ccccccc}
\hline
Molecule & Band & Line & $\lambda$$_{air}$(\AA~) & W$_{\lambda}$(m\AA~) & N(J$^{"}$) (10$^{13}$cm$^{-2}$) & \\

\hline 
C2 & A-X (3-0) & R(0) & 7719.33 & 3(1)&0.91 (0.3)\\
&&R(2) & 7716.53 & 5(2)&3.56 (1.4)&\\
&&R(6)+R(4) & 7714.58+7714.95 & 5(2)$^{a}$+3(2)$^{a}$&1.98(0.8)+2.78 (1.8)&\\
&&R(10) & 7717.47 & 1.5(1)&1.6 (1)&\\
&&Q(2) & 7722.10 & 4.3(1)&2.04 (0.5)&\\
&&Q(4) & 7724.22 & 6.5(1)&2.69 (0.5)&\\
&&Q(6) & 7727.56 & 3(1)&1.45 (0.5)&\\
C2 & A-X (2-0) & R(6) & 8750.85 & 4.3(0.5)&1.51 (0.4)&\\
&&R(8)+R(4) & 8751.49+8751.68 & 3(1)$^{a}$+6.5(1)$^{a}$&1.11 (0.4) +2.13 (0.3)&\\
&&R(10)+R(2) & 8753.58+8753.95 &2.4(1)$^{a}$+14.8(1)$^{a}$&0.91(0.4)+4.04 (0.6)&\\
&&R(0) & 8757.69 & 8.4(1)&0.91 (0.1) &\\
&&Q(2) & 8761.19 & 10(1)&2.16 (0.2)&\\
&&Q(4) & 8763.75 & 11.9(1)&2.57 (0.2)&\\
&&P(2) & 8766.03 & 3$^{a}$& &\\
&&Q(6) & 8767.76 & 7.4$^{a}$&1.60 (0.4)&\\
&&Q(8)+P(4) & 8773.22+8773.43& 4(2)$^{a}$+5(2)$^{a}$&1.29(0.6)+3.00(0.5)&\\
&&Q(10) & 8780.14 & 2.5(1)&0.54 (0.2)&\\
&&Q(12) & 8788.56 & 3 (1) &0.64 (0.2)&\\
\hline
\end{tabular}
}
\end{center}
\end{minipage}
\end{table*}

In order to obtain average column densities $<N(J")>$ in rotational level J", the column densities inferred from
the individual measurements in the R and Q lines of the (2,0) and (3,0) bands, when available, were combined. Following van Dishoeck and Black (1982), we obtain the  C$_2$ excitation diagram for Cernis 52 shown in Fig. 3 where we represent the weighted relative column densities  -ln [5 N (J$^{"}$) / ((2J$^{"}$+1) N(2))]  versus excitation energy  E(J") of rotational level J$^ {"}$ (k is the Boltzmann constant).  Rotation excitation temperatures  were obtained from a linear fit to the logarithm of the column densities of the two, three and four lowest rotational levels starting from J"=0 and   result   T$_{02}$=34 K,  T$_{04}$=39 K and  T$_{06}$=49 K, respectively.

\begin{figure*}
\includegraphics[angle=0,width=11cm,height=11cm]{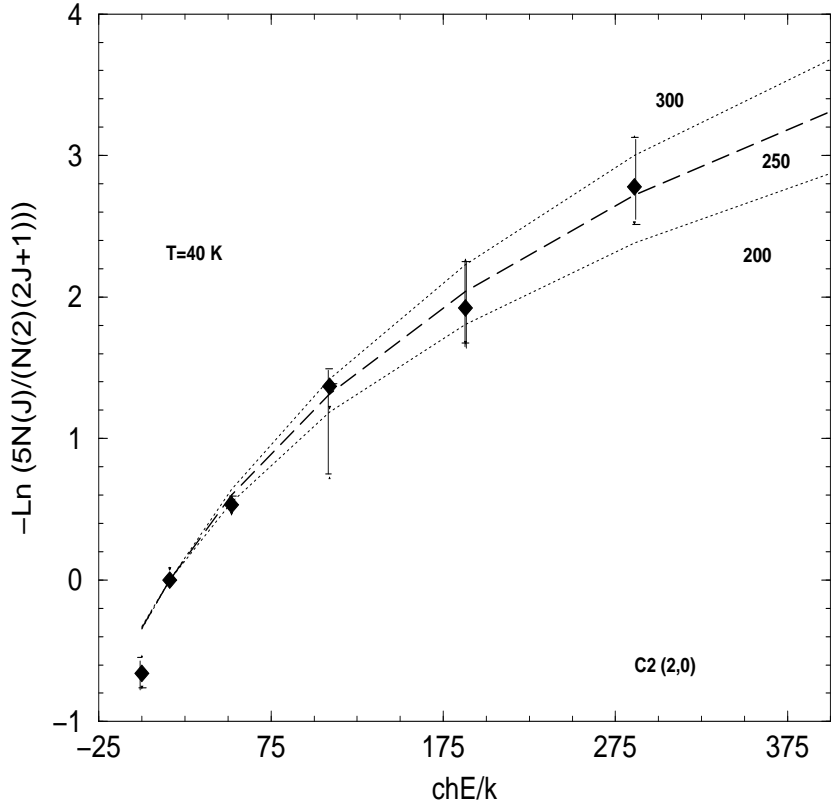}
\caption{C$_2$ excitation diagram for Cernis 52, with observed relative rotational populations with respect to that of the $J^{"}$=2 level as function of the excitation energy E($J^{"}$). Filled diamonds correspond to individual line detections. The 1$\sigma$  error bars are indicated. The theoretical populations  (dotted and dashed lines) at a kinetic temperature T = 34 K are shown for comparison at several densities n (see text). The dot-dashed line indicates the thermal distribution at T=34 K.}
\label{fig:fig4}  
\end{figure*}

  As described by van Dishoeck and Black (1982), in general  the populations of all rotational levels  cannot be characterized by a single rotational temperature,  the detailed behaviour of the rotational excitation temperature depends
  on the gas kinetic temperature T$_{kin}$, on the intensity of  the interstellar radiation field I and on the collisional rate n$\sigma_0$ (the  number density of collision partners n= n(H) + n(H$_2$) times the cross section,
  $\sigma_0$,  for collision induced transitions for level J" to rotational level (J" - 2)). To interpret the excitation diagram of Fig. 3 we assume the same  $\sigma_0$=2 x 10$^{-16}$ cm$^{2}$ and the same  intensity of the
  interstellar radiation field than van Dishoeck and Black (1982) and adjusted for the differences with respect their adopted $f$ values. We infer from these models  an  excitation temperatute of 34$\pm$10 K and for the effective
  density of collision partners n=300 $\pm$ 80 cm$^{-3}$ (see Fig. 3). We checked this result using the code made available by B. Mc Call at  the web site  http:/dib.uiuc.edu/c2/.  We generated a grid of models for  gas kinetic temperatures ranging from 10 to 1000 K, with a step of 1 K and a range of particle densities from n=50 to 2000 cm$^{-3}$ with a step of 50 cm$^{-3}$. The results were scaled to the f-values adopted above (a factor 1.36 higher than  in the code by Mc Call) for the (2,0) series and we took for comparison the weighted mean of the column densities N(J)  inferred from the (2,0) series of lines.  Then, we determined from the best figure of merit (obtained from the minimum difference between predicted and  observed values) the most likely values for T and n. The best fit was found at  T=40$\pm$10 K and n=250$\pm$ 50 cm$^{-3}$ in very good agreement with our previous independent analysis.

Total observed column densities were derived from the sum $\Sigma_{J^{"}}$  N(J") over the observed rotational levels of each individual band, resulting N$_{obs}$= (9.2$\pm$1.4) x 10$^{13}$ cm$^{-2}$ and (9.8$\pm$1.5) x 10$^{13}$ cm$^{-2}$ for the (2,0) and (3,0) bands, respectively. The combination leads to  N$_{obs}$=(9.5$\pm$1.2) x 10$^{13}$ cm$^{-2}$.  The total C$_2$ column density, defined as the sum of the mean column densities of the observed levels and of the contribution of the unobserved levels estimated from the theoretical model characterized by the best-fitting parameters results N$_{tot}$= (10.5$\pm$1.2) x  10$^{13}$ cm$^{-2}$.

Heliocentric velocities for individual   interstellar absorption lines of C$_2$ were determined from the final combined spectrum obtained each of the four  days that Cernis 52 was observed.
Typically, the three or four most intense unblended lines for each band were used. We found 14.9$\pm$1.0  and 12.7$\pm$1.2 km s$^{-1}$ as average values of the heliocentric velocities of the lines
measured for the (2,0) and (3,0) bands, respectively.  The  error was obtained as the rms of the measurements. We find consistent results among the two bands and finally combine them to obtain an 
average heliocentric velocity of  v$_{hel}$=13.8$\pm$0.9 km s$^{-1}$ as  best estimate from  the detected interstellar C$_2$ lines.  It appears,  at this spectral resolution,  that C$_2$ lines
rise in a single velocity component.    The C$_2$ lines and the K I line (see below) display the same heliocentric velocity within the uncertainties of our measurements.

\subsection{Interstellar Potassium}

The K I 7664 and 7698 \AA~ absorption lines are also present in our spectra of Cernis 52. The K I 7698 line is well separated from a telluric O$_2$ absorption line but the K I 7664  line is not. The equivalent width of the K I 7698 A line is 153$\pm$3 m\AA. We have obtained heliocentric velocities for  this K I line  from the available set of  spectra  resulting a final mean value of 13.7 $\pm$ 0.4  km s$^{-1}$ where the error is obtained from the dispersion of the measurements obtained in different days. The heliocentric velocity  agrees well with those of C$_2$ above  and CH, CH$^+$ measured by Iglesias-Groth et al. (2010). 
Using the atomic parameters of Morton (1991), we infer N(K) = 1.4  x 10$^{12}$ cm$^{-2}$ towards Cernis 52.

\section{Discussion}

  Out of 37 lines of sight towards translucent molecular clouds  investigated by van Dishoeck and Black  (1989, see their table 4) 19 present E(B-V) higher than Cernis 52 but only 4 show C$_2$ column densities higher than this star. Remarkably,  the column density of Cernis 52 is  about 2 times larger than the average value observed toward stars with similar E(B-V). Out of 12 lines of sight investigated by Gredel (1999) with E(B-V) higher than Cernis 52 only 2 showed C$_2$ column densities comparable or higher than Cernis 52. The C$_2$ column density derived for Cernis 52 is one of  the  largest so far observed in translucent molecular clouds.  Although its value is not as large as the one toward the much more heavily reddened star Cygnus OB2 No. 12 with E(B-V)=3.31 reported to be 20 x 10$^{13}$ cm$^{-2}$ (Gredel et al.  2001).   It is however comparable to the highest values reported for the stars in  Gredel (1999) where star HD 172028  displays one of the highest  C$_2$ column densities (N(C$_2$)=  (11$\pm$0.5) x  10$^{13}$ cm$^{-2}$) comparable to  our measurement for Cernis 52. HD 172028 displays moderate extinction (A$_V$=2.3) as and Gredel determines a kinetic temperature of 30 K and density of 300$\pm$100 cm$^{-3}$ for the intervening cloud, both values are  very similar to those found  for the cloud  in the line of sight of Cernis 52.

 The heliocentric velocities of C$_2$ and CH in Cernis 52 coincide so it is  very likely that both molecules co-exist to a large extent in the same region. The C$_2$ and CH column densities of translucent molecular clouds are empirically well correlated and appear to  follow the relationship log N(C$_2$) = 0.85 log N(CH)+2.2 ( van Dishoeck \& Black 1989). Theoretical reasons for this correlation are  discussed by Federman et al. (1994). Using the previous  relationship and  the CH column density N(CH) = 20 ($\pm$ 2) x 10$^{13}$ cm$^{-2}$ of Iglesias-Groth et al. (2010) we expected  a column density of C$_2$ of  $\sim$ 22 x 10$^{13}$ cm$^{-2}$. This is   about a factor 2 higher than  our measurement, however the column density  of CH derived for Cernis 52 is significantly affected by saturation of the observed absorption line and the   applied correction may have led to an  overestimation of the abundance by some 50\%. With this consideration in mind and using  the relation N(H$_2$) = 2.6 x 10$^7$ N(CH)  of Somerville \& Smith (1989) we estimate  a column density of  N(H$_2$)= 5  x 10$^{21}$  cm$^{-2}$, but cannot discard a value a  factor two lower because of the saturation correction just mentioned.  The fractional abundance  f(C$_2$) = N(C$_2$)/N(H), that is f(C$_2$)$\sim$ N(C$_2$)/2N(H$_2$) $\sim$ N(C$_2$)/ 2 N(H) is  $\sim$ 1 x 10$^{-8}$ which compares well with  values 0.5-1.5 x 10$^{-8}$ obtained by Gredel (1999) toward stars in the assocation surrounding  NGC 2439, in  Vela OB 1  and in Cen OB1 in spite of generally higher gas kinetic temperatures  and densities towards these lines of sight. 

Observations of the $A ^1$$\Pi_u$ - $X ^1$$\Sigma^+$$_g$ transition of C$_3$ at 4051.6 \AA~ in translucent sight lines have been reported by Oka et al. (2003) and \'Ad\'amkovics et al. (2003). The observed C$_3$ column densities range from 10$^{12}$ to 10$^{13}$ cm$^{-2}$ and are well correlated with the corresponding C$_2$ columns with a ratio  N(C$_2$)/N(C$_3$)$\sim$40. The observed strong correlation suggests that C$_3$ and C$_2$ are involved in the same chain of chemical reactions.  As discussed by Oka et al., C$_3$ formation may result from C$_2$ by photoionization, ion neutral reactions and dissociative recombinations. The spectral range of our  HET spectra do not cover this transition, but fortunately, the spectra of Cernis 52   obtained with the blue arm of  ISIS at the 4.2m William Herschel Telescope (described in Iglesias-Groth et al. 2010) do include the relevant spectral range to search for this transition.   We report the detection of an absorption feature with an equivalent widht of W$_{\lambda}$= 6$\pm$1.5  m\AA~   fully consistent in wavelength with the transition from C$_3$.  The feature presents a weaker blend on the blue wing whose origin can be understood looking at the simulated spectra by Haffner and Meyer (1995) for the 4051.6 band of C$_3$ in clouds of T=40  K which fits well the conditions of our cloud, or simulations for  a larger range of temperatures by \'Ad\'amkovics et al. (2003). The blue line of the blend appears to be   caused by the pileup of robovibronic transitions of the first  lines of the R series in the wavelength range 4050-4051 \AA. The main feature is due to the stronger lines of the Q  series in the range 4051-4052 \AA. Our  equivalent width  would mostly correspond to the pileup of these Q-branch lines. The resolving power of the WHT spectra is not sufficient to measure individual rotational lines and therefore does not allow us to derive the temperature of the cloud from C$_3$ which would have provided an interesting comparison with the results  derived from C$_2$.  Our data  provides however a preliminary estimate of the column density for C$_3$. Adopting the simplest assumption of a one-temperature rotational distribution, following  the  arguments given by Oka et al. (2003) (we multiply  by a factor 2 the right term of the W$_{\lambda}$ - N expression in  Section 3.1) and using  their oscillator strength of  f=0.016   we derive N(C$_3$)= (5.2$\pm$1.3) x 10$^{12}$ cm$^{-2}$ and a ratio of N(C$_2$) / N(C$_3$)= 20$^{+10}_{-6}$ which is comparable to those derived in the above papers for other lines of sight.

 Models of quiescent translucent molecular clouds are presented by van Dishoeck and Black (1988) for a range
of total visual extinction A$_V$ (1-5 mag), densities (500-1000 cm$^{-3}$) and temperatures (15-40 K). Computed column densities for several species (CH, C$_2$, CN, CO, etc...) in these models can also  be found in van Dishoeck \& {Black (1988) which may help to predict the abundance  of other species in the intervening cloud and prepare observing programmes for an extensive characterization. The  constraints on density (250$\pm$50 cm$^{-3}$) and gas kinetic temperature
 (40$\pm$10 K)  that we have obtained provide already valuable information for the modelling of the anomalous microwave emission in the Perseus complex. The models computed  by Iglesias-Groth (2005) to explain the spectral energy distribution of electric dipole radiation from hydrogenated carbon particles in this light of sight  should be revised accordingly. However, we recall here that to infer the density of the cloud  a major assumption was made on the
value of the intensity of the radiation field in the near infrared, and since the excitation temperature curves are degeneratedversus  the ratio of the density to the intensity field, we would still require an independent determination of the radiation field instead of an assumption on its value,  to determine proper abundances. For instance, the strength of the radiation field was inferred to be enhanced by a factor of 3-5  in  $\zeta$ Oph and other diffuse clouds with similar kinetic temperatures  (van Dishoeck \& Black 1988). The incident radiation field in  the translucent cloud  may also be enhanced because of the presence of Cernis 52.  For the moment, with the limited number of molecular species so far observed toward Cernis 52, we cannot constrain the parameter of the radiation field any further.   However, plausible  models for the Cernis 52 cloud  predict significant column densities for a variety of atomic and molecular species which may be observed in the near future  (H, H${_3^+}$, C$_2$H, C$_2$H$_2$, OH, H$_2$O, NH, CH$_2$, etc.)  and could in principle provide much better constrains on the basic parameters of the model. Simple steady state chemical kinetics (Oka et al.) indicate that the neutral molecules C$_2$, C$_2$H, C$_2$H$_2$ and C$_3$ are more abundant than the ionic species by at least 2 orders of magnitude.
 
\subsection{Comparison with other lines of sight in Perseus}

The diffuse clouds toward $\zeta$ Per and $omi$ Per  have been extensively studied since early work  on interstellar C$_2$ (Black, Hartquist and Dalgarno 1978, Hobbs 1979, Chaffee et al. 1980, van Dishoeck \& Black 1989).   These two stars are located in the Perseus molecular complex, but in  lines of sight where anomalous microwave emission is not detected, if this emission exists it is definitely much weaker than in the line of sight of Cernis 52.   According to van Dishoeck \& Black (1982) the molecular core of the $\zeta$ Per cloud would have n= 150cm$^{-3}$, T= 45 K and I$\sim$=0.7 and the column density of C$_2$ would be N(C$_2$)= 1.6 x 10$^{13}$ cm$^{-2}$ which compares well with the predictions in the early model  by Black et al. (1978). A recent summary of results on gas kinetic temperatures and densities for these and other  diffuse and translucent clouds can be found in  Sonnentrucker et al. (2007).  The physical conditions in these two clouds are apparently  not very different to those estimated for   Cernis 52 but the extinciton is 3 times higher  and  
the abundance of C$_2$ and CH  are $\sim$10 times higher than in $\zeta$ and $omi$ Per. It is possible that reactions causing destruction (C$_2$  + H$_2$ $\rightarrow$ C$_2$H + H) and production (C$_2$H + h$\nu$ $\rightarrow$  C$_2$ + H)  of C$_2$ work at very different rates at the lower temperature of the Cernis 52 cloud. A  search for  lines of C$_2$H  in the optical or radio would provide valuable insight on the reasons for the much higher column density of C$_2$ in Cernis 52 and will contribute to a better understanding of the physical conditions there and in particular of the processes involved in the formation of  naphthalene and anthracene cations in the intervening cloud (Iglesias-Groth et al. 2008, 2010).

\section{Conclusions}

We have presented observations of interstellar absorption lines of C$_2$ toward Cernis 52. We have identified six rotational lines of the (2,0) and five rotational lines of the (3,0)  Phillips system. The population distribution in the lowest rotational levels  indicates a rotational excitation temperature of T$_{rot}$ = 34 $\pm$ 12 K. A detailed study of the level population leads to a best fit  gas  kinetic temperature T = 40$\pm$10 K and particle density n = 250 $\pm$50  cm$^{-3}$ for the molecular gas in the intervening cloud. We obtain a total C$_2$ column density of (10.5$\pm$1.2) x 10$^{13}$ cm$^{-2}$. We also report a tentative detection of an interstellar absorption feature at the wavelength (4051.6 \AA) expected for an unresolved blend of the strongest Q-branch lines arising from the ground vibrational state of the C$_3$ $A ^1$$\Pi_u$ - $X ^1$$\Sigma^+$$_g$ electronic transition and derive from this  a column density for C$_3$ of (5.2$\pm$1.3) x 10$^{12}$   cm$^{-2}$. The ratio C$_2$/C$_3$ compares well with those reported in the literature on translucent clouds (Oka et al. 2003). Higher resolution, higher S/N data will be required to confirm the C$_3$ detection and to determine the weak rotational structure of the band which  could be used as an independent diagnostic of the cloud's kinetic gas temperature. In addition, it is worth  to search for longer multicarbon chains  (C$_4$, C$_5$,... C$_{18}$) in this rather carbon-rich sight line.  The growth of carbon chains from C$_2$ and C$_3$ via C$^+$ association followed by H abstraction and recombination may be fast  after C$_4$ (Freed et al. 1982). This line of sight is well suited to carry out such studies which will provide insight  on carbon chemistry.  

In the intervening cloud toward Cernis 52, C$_2$ forms dominantly in cool material at gas kinetic temperatures of 40 $\pm$ 10 K. The C$_2$ and CH column densities are apparently  both high   and their relative value follow the  expectations from theoretical models of translucent molecular clouds and the well established empirical relationship for this two species.  The C$_2$ molecules mainly probe the dense cold core of the cloud,  measurements of column densities of H, H$_2$ in various rotational levels and C in various fine structure levels would be worth to  model in detail  the intervening cloud  and gain insight on the physical processes causing the  anomalous microwave emission, it is important to obtain also information on a possible warmer less dense envelope. Observations of C$_2$H may help to explain the high content of C$_2$. This molecule  may provide pathways to the formation of long-chain carbon molecules and is likely that a high content may have also favoured the formation of  PAHs to a detectable level. Additional observations may establish the chemistry of the cloud and  the role that a high content of C$_2$ and CH could have  on  the formation  of PAHs. 
 
\section*{ACKNOWLEDGEMENTS} 
I thank  D.L. Lambert and R. Rebolo for useful comments on this paper. This work was partially supported by  grant AYA-2007-64748 from the Spanish Ministry of Science and Innovation.

\bibliographystyle{mn2e}

\end{document}